\documentclass[aps,manuscript,showpacs,psfig]{revtex4}                                                          

\usepackage{graphicx}
\usepackage{psfig}

\begin{document}

\baselineskip 0.8cm

\title{Probing gluon helicity distribution and quark transversity through
hyperon polarization 
in singly polarized $pp$ collisions}
\author{Xu Qing-hua, 
 and Liang Zuo-tang}
\affiliation{
Department of Physics, Shandong University,
Jinan, Shandong 250100, China}


\begin{abstract}

We study the polarization of hyperon in different processes in singly
polarized $pp$ collisions, in particular its
relation to the polarized parton distributions.
We show that by measuring hyperon polarization in 
particularly chosen processes, one can extract useful information on these
parton distributions. We show in particular that, by measuring the 
$\Sigma^+$ polarization in high $p_T$ direct
photon production process, one can extract information on the gluon helicity
distribution; and by
measuring the transverse polarization of hyeprons with high $p_T$ 
in singly polarized reactions, one can obtain
useful information on the transversity distribution.
We present the 
numerical results obtained for those hyperon polarizations using 
different models for parton distribution function and those for
the spin transfer in fragmentation processes.

\end{abstract}

\pacs{13.88.+e, 13.85.Ni, 13.87.Fh}

\maketitle

\newpage

\section{Introduction}

The spin structure of nucleon is one of the important issues in high energy spin physics. 
Since the ``spin crisis'' induced by the EMC data in 1988 \cite{EMC88}, 
the study of it has been an active area of both experimental and
theoretical research.
Presently, the helicity distributions of quarks, in particular the valence quarks, 
have been determined with relatively high precision experimentally through inclusive 
deeply inelastic lepton-nucleon scattering (DIS) processes.
However, 
little is known from experiment about the gluon helicity distribution $\Delta g(x,Q^2)$,
except some indirect estimates from the $Q^2$ evolution of the quark
distributions \cite{Ji01}.
For the transversely polarized case, i.e., the transversity distribution of
quarks in the nucleon, $\delta q(x,Q^2)$, the situation is even worse.
Due to its chiral-odd property, it can not be measured in
inclusive DIS process and is up to now unmeasured \cite{trans02}.
It has to be accompanied by another chiral-odd quantity.
Excitingly, many interesting ideas to measure $\Delta g(x,Q^2)$ and $\delta
q(x,Q^2)$ in polarized $ep$ scatterings and $pp$ collisions
have been proposed \cite{RHIC,trans02,Ji01}
and the related experiments are being undertaken or under way.

Hyperon polarization, in particular for $\Lambda$,
has been widely used to study various aspects of spin effects
in different reactions in particular
the spin-dependent fragmentation functions for their self-analyzing decay.
The spin-dependent fragmentation function is interesting in itself
and also in serving as filters for exotic parton distribution functions,
such as polarized antiquark distributions and the above-mentioned gluon
helicity distribution and transversity of nucleon \cite{Jaffe01}.

In a recent publication \cite{XLL02}, we made a detailed study of 
the longitudinal polarization of hyperons with high transverse momenta $p_T$ in inclusive
process in $pp$ collisions with one beam longitudinally polarized. 
We found out in particular that the origin of $\Lambda$ is usually very complicated and
the contribution of decay of other hyperons is very large. 
On the contrary, the origin of $\Sigma^+$ is much cleaner.
The contribution from decay of other hyperons is very small.
In addition,
there is a characteristic feature for $\Sigma^+$ production with high $p_T$ in $pp$
collisions, i.e.,
the contributions from those which are directly produced and contain a
fragmenting $u$ quark from the hard subprocess play the dominant role.
In particular for the region of $\eta>1.5$ where $\eta$ is the pseudo-rapidity of
produced hyperons, such contributions are larger than 93\%  
at $p_T>13$ GeV and $\sqrt s=500$ GeV(c.f. Fig.10 of \cite{XLL02}).

This characteristics of $\Sigma^+$ production is remarkable, 
since in the limiting case that
only this kind of contribution is considered, 
the $\Sigma^+$ polarization is 
$P_L^{\Sigma^+(lim)}=t^F_{\Sigma^+,u} P^u_L$,
(where $P^u_L$ is the longitudinal
polarization of the $u$ quark after the hard scattering and
$t^F_{\Sigma^+,u}$ is the spin transfer factor in the fragmentation
process from the $u$ quark to $\Sigma^+$ for the case that $\Sigma^+$
contains the $u$ quark).
There are two different models for the spin transfer in the fragmentation
processes, one is based on the SU(6) wave-function of the hadrons, 
the other is based on the DIS data and those of hyperon decay.
The spin transfer factor
is different in the two pictures but is a constant
in both cases.
Hence, $P_L^{\Sigma^+}$ is directly proportional to $P^u_L$.
The fragmentation effects come in only through the proportional constant
$t^F_{\Sigma^+,u}$.
In this case, measuring $\Sigma^+$ polarization
provides a nice tool to study the polarization of the quarks before
fragmentation.
Since such quark polarization is determined by the polarized quark
distribution functions in the nucleon and the calculable spin transfer
factor in the hard scattering, we expect that we can use it
to study the exotic distributions, i.e.,
the gluon helicity distribution and 
the quark transversity distribution in suitable processes.

In this paper, after a brief summary of the calculation method 
and main conclusion for hyperon polarization in inclusive production
processes in $pp$ collisions,
we present two examples in this direction, i.e.,
hyperon polarization in direct photon production process
in longitudinally polarized $pp$ collisions, and the transverse polarization
of hyperons with high $p_T$ in transversely polarized $pp$ collisions. 
We show that they can be used to study the gluon helicity distribution
$\Delta g(x,Q^2)$ and the quark transversity distribution 
$\delta q(x,Q^2)$ respectively.
We present the numerical results for the hyperon polarization in these
two processes obtained by using different models for $\Delta g(x,Q^2)$
and $\delta q(x,Q^2)$. We show that
both of them can be measured at e.g., BNL RHIC and such measurements
should provide important
information for $\Delta g(x,Q^2)$ and $\delta q(x,Q^2)$.

\section{Calculation method of polarization of hyperons with high $p_T$ in
polarized $pp$ collisions}

The calculation method of the polarization of hyperons with high $p_T$
in longitudinally polarized $pp$ collisions has been given in \cite{XLL02}.
The method can be easily extended to other cases. 
In this section, we summarize the main points of the method.

\subsection {General formulae}

We consider inclusive hyperon production with high $p_T$ in $pp$ collisions
with one beam longitudinally polarized.
The produced hyperons with high $p_T$ are mainly
from the fragmentation of a scattered high $p_T$ parton in the
hard subprocess,
in which one of the initial partons is (longitudinally) polarized.
Since the polarization can be transferred to the outgoing parton in the hard
scattering, the scattered parton can also be polarized and
its polarization can further be transferred to the produced hyperons.
Their polarizations can be obtained as
follows (the subscripts ``+" and ``$-$" below denote helicities) :
\begin{equation}
P_L^{H}= \frac
{d\sigma^{(p_+p \to  H_+X)}-d\sigma^{(p_+p \to  H_-X)}}
{d\sigma^{(p_+p \to  H_+X)}+d\sigma^{(p_+p \to  H_-X)}}
\equiv \frac
{d\Delta \sigma^{(\vec pp \to  \vec HX)}/d\eta}
{d\sigma^{(pp \to  HX)}/d\eta},
\label{gener1}
\end{equation}
where $\eta$ is the pseudo-rapidity of produced hyperon,
$d\Delta \sigma$ and $d\sigma$ are polarized and unpolarized cross
sections for inclusive hyperon production with high $p_T$.
For sufficiently high $p_T$, the factorization
theorem can be applied and the (polarized) cross section can be expressed as a
convolution of perturbatively calculable partonic cross
sections with certain sets of (polarized) parton distribution and
fragmentation functions at a proper scale. In this case the relevant
polarized cross section can be expressed as \cite{Factorm,Florian98},
\begin{equation}
\frac {d\Delta \sigma^{(\vec pp \to  \vec HX)}}{d\eta}
=\int_{p_T^{min}}dp_T
\sum_{abcd}\int dx_a dx_bdz \Delta f_a(x_a,\mu^2)f_b(x_b,\mu^2)
 \Delta D_c^H(z,\mu^2)
 \frac {d\Delta \hat {\sigma}^{(\vec a b\to \vec c d)}}
{d\eta}
\label{desig2}
\end{equation}
where the sum runs over all possible subprocesses, and 
the transverse momenta $p_T$ of hyperons are
integrated above a lower limit $p_T^{min}$;
$\Delta f_a(x_a,\mu^2)$ and $f_b(x_b,\mu^2)$ are the longitudinally
polarized and unpolarized distribution functions of partons in proton
at the scale $\mu$;
$x_a$ and $x_b$ are the corresponding momentum fractions carried
by parton $a$ and $b$.
$\Delta D_c^H(z,\mu ^2)$ is the longitudinally polarized fragmentation
function,
\begin{equation}
\Delta D_c^H(z,\mu ^2) \equiv
D_{c(+)}^{H(+)}(z,\mu ^2)-D_{c(+)}^{H(-)}(z,\mu ^2),
\end{equation}
where $D_{c(+)}^{H(+)}(z,\mu ^2)$ and $D_{c(+)}^{H(-)}(z,\mu^2)$
are the probabilities to produce a $H$ with positive and negative helicity
in the fragmentation of a parton $c$ with positive helicity, carrying off a fraction $z$
of parent parton's momentum.
$d\Delta \hat {\sigma}$ is defined for the hard scattering
subprocess similarly as $d\Delta {\sigma}$ in Eq.(\ref{gener1}),
which can be calculated by pQCD.
                                                                                                                
The unpolarized cross section $d\sigma^{(pp\to HX)}/d\eta$
is given by an expression similar to that in Eq.(\ref{desig2}),
with all $\Delta$'s removed.
It can be calculated with considerable precision using
the available parameterizations of the
unpolarized parton distribution and fragmentation functions.

For the polarized case, we see that from Eq.(\ref{desig2}),
the unknown parts in the formulae are the polarized fragmentation
function $\Delta D_c^H(z,\mu ^2)$ and the polarized parton
distribution $\Delta f_a(x_a,\mu^2)$.
If we know $\Delta f_a(x_a,\mu^2)$, we can study $\Delta
D_c^H(z,\mu ^2)$ through hyperon polarization.
Similarly, if the influence from $\Delta D_c^H(z,\mu ^2)$ can be determined,
we can use it to
study the polarized parton distribution functions.

\subsection{The polarized fragmentation function $\Delta D^H_c(z,\mu^2)$}

We consider a general fragmentation process $ q_f\to H_i+X$
(subscripts ``$f$" and ``$i$" denote the quark flavor and the type of hyperons)
and calculate $\Delta D^{H_i}_{q_f}(z,\mu^2)$ in the following way.
We divide the produced $H_i$'s into four
groups and consider them separately.
(A) directly produced and contain the $q_f$'s;
(B) decay products of heavier
hyperons which were polarized before their decays;
(C) directly produced but do not contain the $q_f$'s;
(D) decay products of heavier hyperons
which were unpolarized before their decays.
In this way, we have,
\begin{equation}
D_{q_f}^{H_i}(z,\mu^2)=
\sum_{\alpha} D_{q_f}^{H_i(\alpha)}(z,\mu^2),
\label{Dfunc}
\end{equation} 
\begin{equation}
\Delta D_{q_f}^{H_i}(z,\mu^2)=
\sum_{\alpha}\Delta D_{q_f}^{H_i(\alpha)}(z,\mu^2),\;\;
\end{equation}
where $D_{q_f}^{H_i(\alpha)}$ and $\Delta D_{q_f}^{H_i(\alpha)}$
are the unpolarized and polarized fragmentation functions for 
hyperons of group ($\alpha$), and $\alpha=A, B, C$ or $D$. 

It is clear that hyperons
from (A) and (B) can be polarized
while those from (C) and (D) are not \cite{GH93,BL98,LL00,LXL2001}.
Hence, if we denote $t^{F(\alpha)}_{H_i,f}$ as the spin transfer factor
for group ($\alpha$) in the fragmentation process, i.e.,
\begin{equation}
t^{F(\alpha)}_{H_i,f}(z,\mu^2) =
\frac{\Delta D_{q_f}^{H_i(\alpha)}(z,\mu^2)} {D_{q_f}^{H_i(\alpha)}(z,\mu^2)},
\label{detf}
\end{equation}
we have,
\begin{equation}
t^{F(C)}_{H_i,f}(z,\mu^2)=t^{F(D)}_{H_i,f}(z,\mu^2)=0.
\end{equation}
                                                                                                                
For those hyperons from group (A),
the spin transfer factor $t^{F(A)}_{H_i,f}(z,\mu^2)$
is taken as the fraction of
spin carried by the $f$-flavor-quark
divided by the number of
valence-quark of flavor $f$ in $H_i$.
In different pictures, such as the SU(6) picture and DIS picture
used in $e^+e^-$ and $ep$ reactions \cite{GH93,BL98,LL00,LXL2001},
the contributions to the hyperon spin from different flavors
are different. 
In the SU(6) picture, these contributions can be obtained
from the SU(6) wave functions of the hyperons.
In the DIS picture, for the $J^P=(1/2)^+$ octet hyperons,
they are obtained from the DIS data on the spin dependent
structure functions and those on hyperon decay.
In both pictures, $t^{F(A)}_{H_i,f}$ is a constant independent of $z$.
Clearly, $t^{F(A)}_{H_i,f}$ is the probability for the polarization of 
$q_f$ to be transferred to the produced hyperon $H_i$ in the case that
$H_i$ contains $q_f$. 
It is called ``fragmentation spin transfer factor" and is usually
denoted as  $t^{F}_{H_i,f}$, i.e.,  $t^{F(A)}_{H_i,f}\equiv t^{F}_{H_i,f}$. 
A list of $t^{F}_{H_i,f}$ for different hyperons
can be found in Table I of Ref. {\cite{LL00}}.
                                                                                                                
For those from group (B), hyperon $H_i$'s are from the decay of
the heavier hyperon $H_j$'s which are from group (A) and polarized
before their decays.
Here, an additional decay polarization transfer factor
$t^D_{H_i,H_j}$ is needed to obtain
$t^{F(B)}_{H_i,f}(z,\mu^2)$ and we have
\begin{equation}
t^{F(B)}_{H_i,f}=t^D_{H_i,H_j}t^{F}_{H_j,f},
\end{equation}
where $t^D_{H_i,H_j}$ is the probability for
the polarization of $H_j$ to be transferred to
$H_i$ in the decay process $H_j\to H_i+X$ and
the superscript $``D"$ stands for decay.
It is determined by the decay process and is independent
of the process in which $H_j$ is produced.
For the octet hyperon decays, they are
extracted from the materials in Review of Particle
Properties \cite{PDG2000}.
For the decuplet hyperons,
we have to use an estimation based on the SU(6) quark model \cite{GH93}.
Thus, as we pointed out in \cite{XLL02}, 
to reduce the uncertainty of calculations, it is important to consider
the hyperons to which decay contributions are small.

Obtaining the spin transfer factors for different groups of hyperons, 
we can get $\Delta D_{q_f}^{H_i}(z,\mu^2)$ if we know the corresponding unpolarized
fragmentation functions $D_{q_f}^{H_i(\alpha)}(z,\mu^2)$.
The result is given by,
\begin{equation}
\Delta D_{q_f}^{H_i}(z,\mu^2)=
t^{F}_{H_i,f}D_{q_f}^{H_i(A)}(z,\mu^2)+
\sum_{H_j}t^D_{H_i,H_j}t^{F}_{H_j,f}D_{q_f}^{H_i(B,H_j)}(z,\mu^2),
\label{DelD}
\end{equation}
where $D_{q_f}^{H_i(B,H_j)}(z,\mu^2)$ is the contribution to $H_i$
production from group (B) through $q_f\to H_j+X$ and $H_j\to H_i+X'$. 

Using Eqs.(\ref{Dfunc}) and (\ref{DelD}), we can obtain the
$H_i$ polarization in the general fragmentation process
$q_f\to H_i+X$. 
We note that, one of the characteristics of the model is that,
the $z$-dependence of $P^{H_i}$ comes from the
interplay of the different contributions $D_{q_f}^{H_i(\alpha)}(z,\mu^2)$,
which are determined by the hadronization mechanism. 
These different contributions can be
calculated using a hadronization model that are well tested
in unpolarized reactions.
This means that the shape of the $z$-dependence of the hyperon polarization
from the fragmentation process $q_f\to H_i+X$ in this model can be fixed
to a very large extent by the results of an unpolarized hadronization
model without any new free parameter.
It is therefore very crucial to test the model by looking at the shape of 
the $z$-dependence of hyperon polarization.
In this sense, the best places to test this model are 
e.g. $e^+e^-$ annihilation or deeply inelastic $ep$ scattering,
where fragmentation of a polarized quark with a given energy can be studied.
The model has been applied to these processes \cite{BL98,LL00,LXL2001}.
Presently, data on $\Lambda$ polarization in $e^+e^-$ annihilation
are available.
It is encouraging to see that the obtained results under both SU(6) and
DIS pictures give a good description of the $z$-dependence of the data.
Further tests can be made by future experiments in
other reactions such as lepton nucleon deep-inelastic scattering 
and $pp$ collisions.

\subsection{Different contributions to hyperon production in  $pp\to H_iX$}

As mentioned earlier, 
the different contributions $D_{q_f}^{H_i(\alpha)}(z,\mu^2)$'s are determined
by the hadronization mechanism, and
are independent of the polarization properties and can be 
obtained from a hadronization model.
Presently, the most convenient way to do this is to employ a Monte Carlo
event generator. 
In practice, we calculate hyperon polarization
by rewriting $P_L^{Hi}$ in Eq.(1) as,
\begin{equation}
P_L^{H_i}={ {\sum\limits_f t^{F}_{H_i,f} P^{q_f}_L \langle n^A_{H_i,f}\rangle
+\sum\limits_{j,f} t^{F(B)}_{H_i,f}P^{q_f}_L \langle n^B_{H_i, H_j, f}\rangle}
 \over
{\langle n^A_{H_i}\rangle +\langle n^B_{H_i}\rangle +
\langle n^C_{H_i}\rangle +\langle n^D_{H_i}\rangle} }.
\label{polh}
\end{equation}
The quantities $P_L^{q_f}$ and the different $\langle
n^\alpha_{H_i}\rangle$'s are related to the parton distribution and the
fragmentation functions in the following way.
$P_L^{q_f}$ is the polarization of the fragmenting quark $q_f$,
i.e., the polarization of the outgoing quark in the hard scattering,
which can be calculated by pQCD and suitable parton distributions.
$\langle n^A_{H_i,f}\rangle$ is the average number of
$H_i$'s which are directly produced and contain $q_f$, and
$\langle n^B_{H_i,H_j,f}\rangle$ is the average number of $H_i$
coming from the decay of $H_j$ which contains a polarized $q_f$.
$\langle n^A_{H_i}\rangle(\equiv \sum\limits_f \langle n^A_{H_i,f}\rangle$),
$\langle n^B_{H_i}\rangle(\equiv \sum\limits_{j,f} \langle
n^B_{H_i,H_j,f}\rangle)$,
$\langle n^C_{H_i}\rangle$, and $\langle n^D_{H_i}\rangle$
are the average numbers of hyperons of group (A), (B), (C),
and (D) in $pp\to H_iX$, respectively.
They are related to $D_{q_f}^{H_i(\alpha)}$ in $pp\to H_i X$ by,
\begin{equation}
\langle n^\alpha_{H_i}\rangle \propto
\sum_{abcd}\int  dx_a dx_bdz f_a(x_a,\mu^2)f_b(x_b,\mu^2)
 D_c^{H_i(\alpha)}(z,\mu^2)
 {d \hat {\sigma}^{( a b\to  c d)}}.
\end{equation}
We use Lund model \cite{lund} implemented by the event generator {\sc
pythia} \cite{PYTHIA}
to obtain them in our calculations for $pp$ collisions.

In Ref.\cite{XLL02}, with the aid of generator {\sc pythia},
we calculated the different contributions to the inclusive high $p_T$
hyperon production in $pp$ collisions. 
We found out that, for $\Lambda$ production (c.f. Fig.7 of Ref.\cite{XLL02}), 
the contribution of group (A) is relatively small. 
The contribution of decay of other hyperons is considerably large.
Hence, the uncertainties in the calculations of
$\Lambda$ polarization are relatively large. 
In contrast, we found out also that, for
$\Sigma^+$ production (c.f. Fig.10 of Ref.\cite{XLL02}),
the decay contribution from heavier
hyperons is very small.
It takes only a few percents for $p_T$$>13$ GeV at $\sqrt{s}$=500 GeV.
The contributions from group (A), i.e., which are directly produced and
contain a fragmenting $u$ quark play the dominant role.
In fact, they give more than 75\%
of the whole produced $\Sigma^+$'s at $p_T$$>$$13$ GeV.
In particular for the region of $\eta$$>$$1.5$, they take more than 93\%.
The reason for this result is simple: The $\Sigma^+$'s of group (A)
are usually the leading particles of quark fragmentation.
They take the largest fractions of the momenta of the fragmenting quark.
To produce a non-leading $\Sigma^+$ with the same $p_T$, 
one needs a quark of much higher $p_T$, whose production is much suppressed.
Since $\Sigma^+$ has two $u$ valence quarks, and $u$-quark contributes
much to the high $p_T$ quark jet, the contribution to $\Sigma^+$
from $u$ dominates at large $p_T$.

If we neglect all other contributions, i.e., consider only the contribution
of type (A), we obtain
the $P_L^{\Sigma^+}$ in this limiting case, i.e.
the $P_L^{\Sigma^+}$ for those only from the origin (A) as,
\begin{equation}
P_L^{\Sigma^+(lim)}=P_L^{\Sigma^+(A)}=
t^F_{\Sigma^+,u} P_L^u.
\end{equation}
It is directly proportional to the polarization
of the $u$-quark, $P_L^u$, after the hard scattering
and the proportional constant is just the
fragmentation spin transfer factor $t^F_{\Sigma^+,u}$.

\begin{figure}
\psfig{file=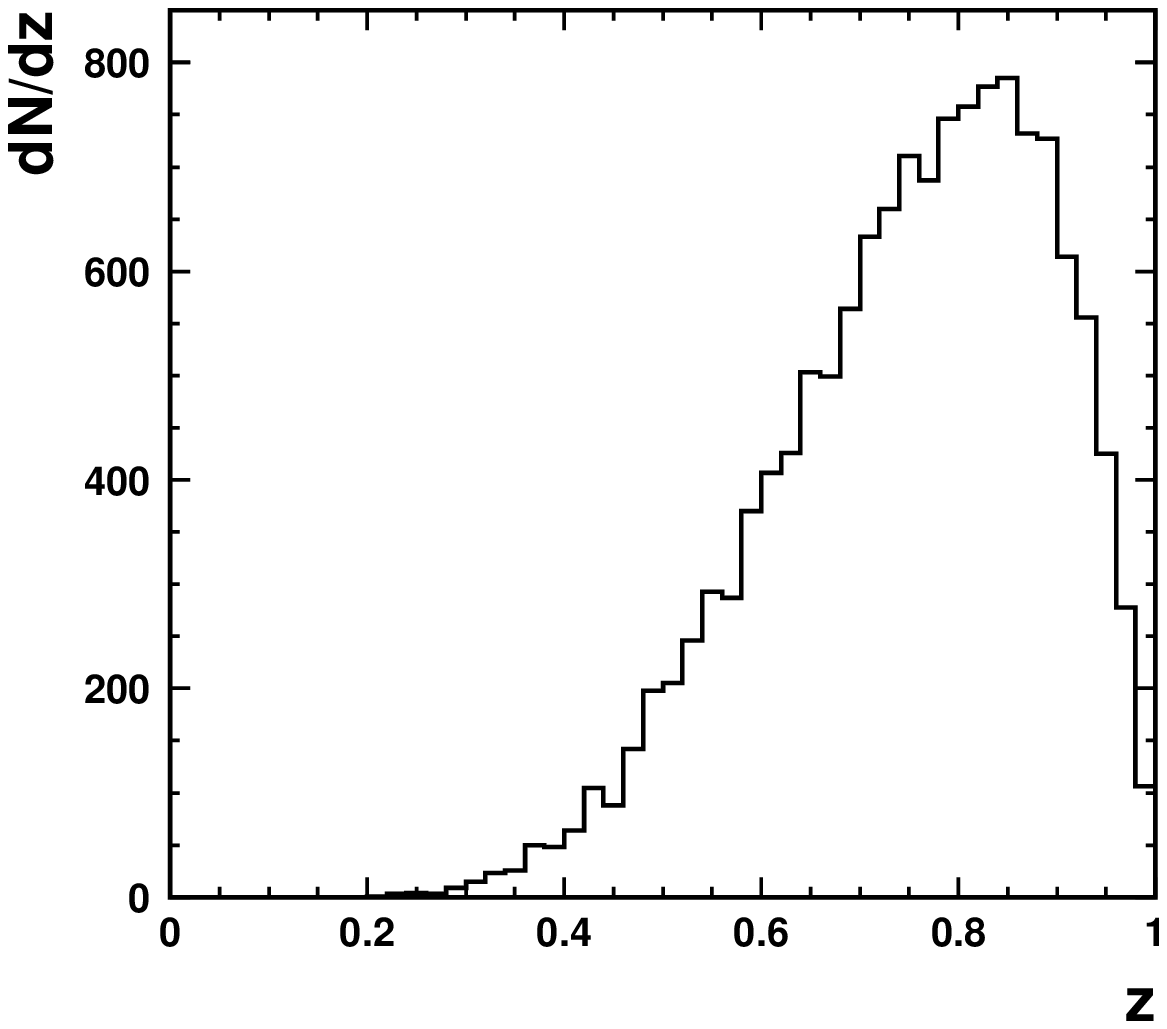,width=16cm}
\caption{The distribution of $z$ fraction of $\Sigma^+$'s that
are directly produced and contain a fragmenting $u$ quark 
in $pp$ collisions at $\sqrt s=500$ GeV and $p_T^{min}=$13 GeV.  
}
\label{ppfr}
\end{figure}

We note that, as can be seen from Eq. (\ref{detf}), 
the factor $t^F_{\Sigma^+,u}$ can in general be $z$-dependent.
But, since the $\Sigma^+$'s from group (A) are usually the leading particles
in the jet, its $z$-distribution is expected to be very narrow.
To get a feeling of it, we calculate the distribution of $z$
for such $\Sigma^+$'s using the MC event generator {\sc pythia} 
and the results are shown in Fig.\ref{ppfr}.
We see that, the distribution is indeed very narrow with a peak at 
$z$$\simeq$0.8.
In this region, the fragmentation function do not change very fast with $z$.
Hence, we expect that
the $z$-dependence of $t^F_{\Sigma^+,u}$, if any,  should not have much
influence on the $\Sigma^+$ polarization.
In this paper, we only discuss the case that such a $z$-dependence is
neglected. In this case,
the factor $t^F_{\Sigma^+,u}$ can be different for different pictures
but it is a constant and can in principle be determined 
in corresponding experiments by one data point.
Hence, $\Sigma^+$ polarization
provides a nice tool to study the polarization of the quarks.
In this way, we can study the polarized parton distributions, such as
the gluon helicity distribution and the quark transversity distribution,
by measuring $\Sigma^+$ polarization in suitable process of $pp$ collisions.
In next two sections, we give two of such examples.

\section{Polarization of $\Sigma^+$ in direct photon production process and
gluon helicity distribution}

Among different methods to probe gluon helicity distribution in nucleon,
high $p_T$ direct photon production in $pp$ collisions
is one of the most direct process 
since the cross section is directly related to the gluon distribution.
The idea to measure gluon helicity
distribution through the double spin asymmetry $A_{LL}$
in direct photon production process is one of the most
promising one in this connection \cite{RHIC}.
Here, both beams need to be polarized. 
In this section, we show that useful information can also be extracted on 
the gluon helicity distribution by measuring polarization of $\Sigma^+$ with
high $p_T$ in $\vec pp \to \gamma {\vec \Sigma}^+ X$ with one beam polarized.

\subsection{Calculation formulae for hyperon polarization in $\vec pp \to
\gamma \vec HX$}

We consider the polarization of hyperons with high $p_T$ 
associated with a single large $p_T$ direct photon in $pp$ collisions
with one proton longitudinally polarized.
We recall that three kinds of hard scattering, i.e.,
$qg\to \gamma q$, $\bar qg\to \gamma \bar q$, and $q\bar q \to \gamma g$,
contribute to $pp \to \gamma HX$ at high $p_T$. 
But, for $pp \to \gamma \Sigma^+ X$, the latter two, i.e.,
$\bar qg\to \gamma \bar q$, and $q\bar q \to \gamma g$,
are significantly suppressed. 
Using the event generator {\sc pythia}, we can study this explicitly.
The results show that, their contributions are less than 4\% totally
at $\sqrt s=200$ GeV and $p_T^{min}=4$ GeV.
This implies that, the hard process $qg\to \gamma q$ play the dominant
role in $pp \to \gamma \Sigma^+ X$ at high $p_T$
and the contributions of other two subprocesses can be neglected.
In this case, the polarized cross section for $pp \to \gamma \Sigma^+ X$
is given by,
\begin{equation}
\frac {d\Delta \sigma^{(\vec pp \to \gamma \vec \Sigma^+ X)}}{d\eta}
=\int_{p_T^{min}}dp_T
\sum_{f}\int dx_a dx_bdz \Delta g(x_a,\mu^2)q_f(x_b,\mu^2)
 \Delta D_f^H(z,\mu^2)
 \frac {d\Delta \hat {\sigma}^{(\vec g q_f\to \gamma \vec q_f)}}
{d\eta}
+(g\leftrightarrow q_f),
\label{dirp}
\end{equation}
where the sum runs over the different quark flavors. 
The first term corresponds to the contribution from 
subprocess $\vec g q_f\to \gamma
\vec q_f$ while
the second term ($g\leftrightarrow q_f$) 
corresponds to the contribution from
$\vec q_fg\to \gamma \vec q_f$, which has a similar
expression as the first term with an exchange of $g$ and $q_f$
in the parton distributions and the hard scattering cross section.
We see that the first term is related to $\Delta g(x)$, 
while the second one is related to $\Delta q_f(x)$.
In our following calculations,
we choose the positive axis of pseudo-rapidity $\eta$ as
the moving direction of the polarized proton.
We expect that the first term dominates at the region of $\eta<0$,
while the second one contributes mainly at the region of $\eta>0$,
since the probability for a forward scattering is usually much larger than
that for a backward scattering.
This will be further illustrated in next subsection using the results from
Monte Carlo calculations.

For the first term of Eq.(\ref{dirp}), the hard scattering is
$ \vec g q_f\to \gamma {\vec q_f}$. The integrand can be rewritten as,
\begin{equation}
\Delta g(x_a,\mu^2)q_f(x_b,\mu^2) \Delta D_f^H(z,\mu^2)
 {d\Delta \hat {\sigma}}
=\sum_{\alpha}P_L^{q_f}t^{F(\alpha)}_{H,f}
g(x_a,\mu^2)q_f(x_b,\mu^2)D_f^{H(\alpha)}(z,\mu^2)
{d \hat {\sigma}},
\label{eqaa}
\end{equation}
where 
$P_L^{q_f}$ is the polarization of outgoing quark in the hard subprocess
$\vec g q_f\to \gamma \vec {q_f}$ and is directly related to $\Delta g(x_a,\mu^2)$ by,
\begin{equation}
P_L^{q_f}=\frac {d\Delta \hat {\sigma}} {d\hat {\sigma}}
\frac {\Delta g(x_a,\mu^2)} {g(x_a,\mu^2)}.
\label{pqT}
\end{equation}
${d\Delta \hat {\sigma}}/{d\hat {\sigma}}$
is the spin transfer from $g$ to $q_f$ in the hard scattering 
$\vec g q_f\to \gamma \vec {q_f}$.
To the leading order, it is given by,
\begin{equation}
D_L(y)\equiv
\frac {d\Delta \hat {\sigma}} {d\hat {\sigma}}
=\frac{1-(1-y)^2}{1+(1-y)^2},
\end{equation}
where $y\equiv k_q\cdot (k_g-k_\gamma)/k_g\cdot k_q$ and
$k_g$, $k_q$ and $k_\gamma$ are the four-momenta
of incoming gluon $g$, quark $q_f$ and the produced $\gamma$ respectively.

For the second term of Eq.(\ref{dirp}), the integrand is,
\begin{equation}
\Delta q_f(x_a,\mu^2)g(x_b,\mu^2) \Delta D_f^H(z,\mu^2)
 {d\Delta \hat {\sigma}}
=\sum_{\alpha}P_L^{q_f}t^{F(\alpha)}_{H,f}
q_f(x_a,\mu^2)g(x_b,\mu^2)D_f^{H(\alpha)}(z,\mu^2)
{d \hat {\sigma}},
\label{eqbb}
\end{equation}
where
\begin{equation}
P_L^{q_f}=
\frac {\Delta q_f(x_a,\mu^2)} {q_f(x_a,\mu^2)},
\end{equation}
since the spin transfer for $\vec q_f g\to \gamma \vec q_f$ is equal to 1.

We further note that
the product on the r.h.s. of Eq.(\ref{eqaa}) or Eq. (\ref{eqbb}),
${g(x_a,\mu^2)} q_f(x_b,\mu^2)
{D_f^{H(\alpha)}(z,\mu^2)} {d\hat {\sigma}}$,
corresponds to the unpolarized hyperon production from group ($\alpha$)
in $pp \to \gamma HX$.
They are proportional to the average numbers $\langle n^\alpha_{H}\rangle$
of hyperon production in $pp \to \gamma HX$, which can also be obtained
by the generator {\sc pythia}.
Hence, if we can determine the constant $t^{F(\alpha)}_{H,f}$ using a few data
points, we will be able to extract information from hyperon polarization 
on the gluon helicity distribution in 
direct photon production process in $pp$ collisions. 

\subsection{Numerical results for $\Sigma^+$ polarization in $\vec pp \to
\gamma {\vec \Sigma}^+ X$}

\begin{figure}
\psfig{file=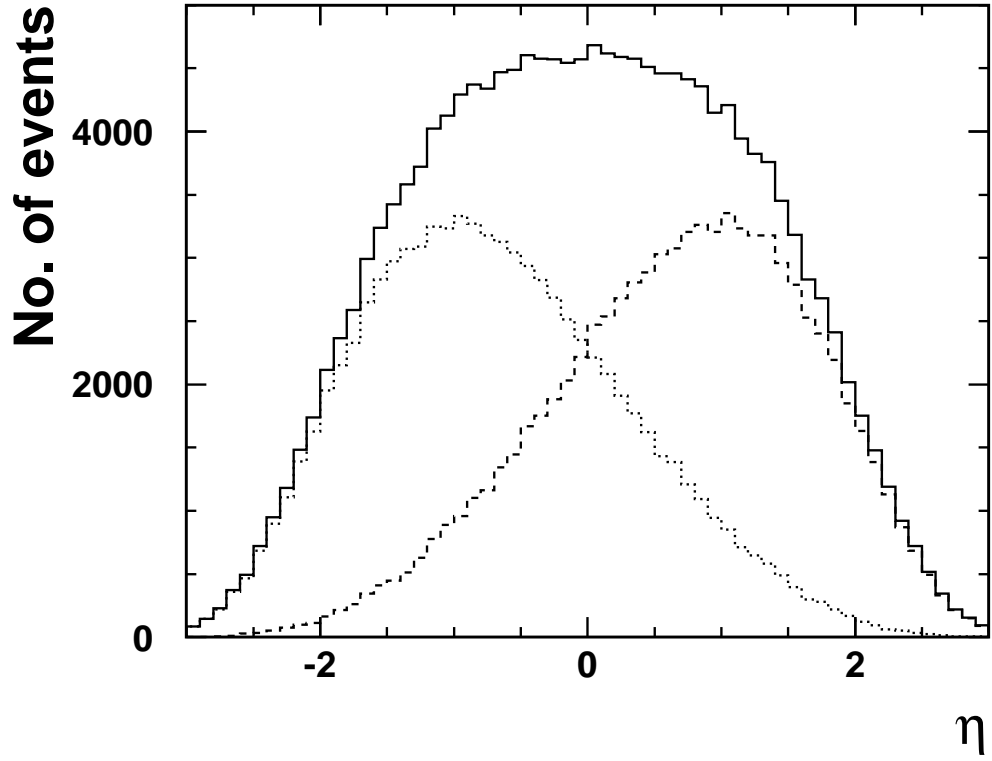,width=16cm}
\caption{Different contributions to $\Sigma^+$ production at
$\sqrt s=200$ GeV and $p_T^{min}=4$ GeV
in the direct photon production in singly polarized $pp$ collisions.
Dotted line denotes the contribution from the hard process
$\vec gq\to \gamma \vec q$, the dashed line denotes that from
$\vec qg\to \gamma \vec q$, and the solid one is the total of them.
}
\label{qgr}
\end{figure}

As can be seen from the above mentioned discussions, to get
information on $\Delta g(x,\mu^2)$,
hyperon production from the subprocess $\vec gq\to \gamma \vec q$
is useful while that from $\vec q g\to \gamma \vec q$ is a background.
Fortunately, the former dominates at the
region of $\eta<0$ while the latter
dominates at the region of $\eta>0$. 
To see this explicitly,
with the aid of generator {\sc pythia}, we calculate
these two contributions separately for $pp\to \gamma \Sigma^+ X$
and the results for 
$\sqrt s=200$ GeV and $p_T^{min}=4$ GeV are shown in Fig.\ref{qgr}.
Here, the cut-off of $p_T$ means that the transverse momenta
of the produced photon and hyperon are both larger than 4 GeV.
We see that although there are some influences
from $\vec qg\to \gamma \vec q$ in the region of $\eta<0$,
they are very small and less than 40\%.
It is possible to use hyperon polarizations in this region 
to study the gluon helicity distribution.

\begin{figure}
\psfig{file=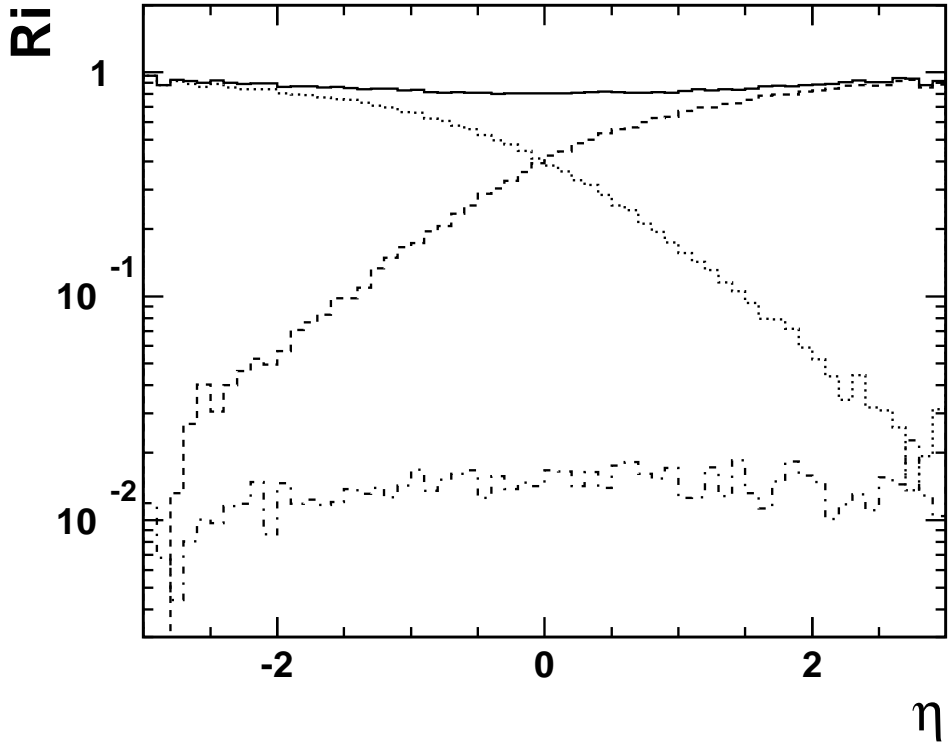,width=16cm}
\caption{Different contributions to $\Sigma^+$ production in
$pp \to \gamma \Sigma^+ X$ as a function of $\eta$
at $\sqrt s=200$ GeV and $p_T^{min}=4$ GeV.
The solid line denotes the contribution of group (A), i.e., those
which are directly produced and contain a fragmenting quark in the 
hard scattering; the dash-dotted line corresponds to the contribution
of decay of other hyperons; the dotted and the dashed lines denote
those which contain a fragmenting quark in the subprocess 
$\vec gq\to \gamma \vec q$ and $\vec qg\to \gamma \vec q$ respectively.
}
\label{sigorg}
\end{figure}

\begin{figure}
\psfig{file=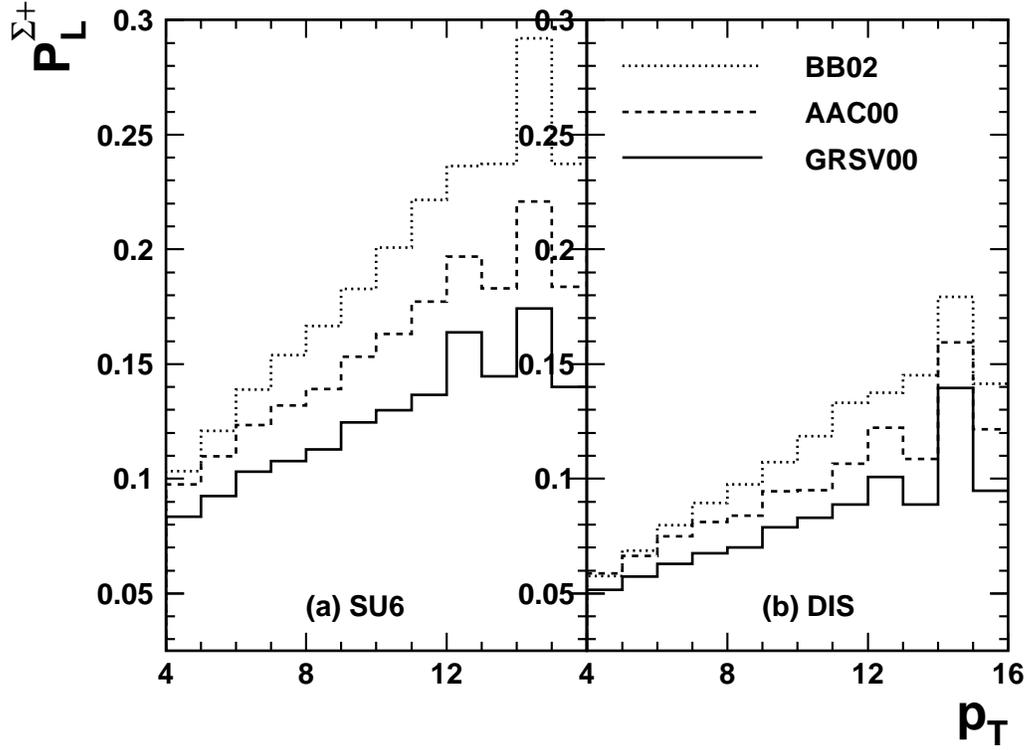,width=16cm}
\caption{Polarization of $\Sigma^+$
for different sets of $\Delta g(x,\mu^2)$
in $\vec pp \to \gamma \vec{\Sigma}^+ X$ as a function of $p_T$
in the region of $-1<\eta<0$ at $\sqrt s=200$ GeV
under (a) SU6 picture and (b) DIS picture for the spin transfer factor.
``GRSV2000" denotes the standard LO GRSV2000 \cite{GRSV2000} parameterization of gluon
distribution;
``BB02" denotes the LO BB02 parameterization\cite{BB02};
``AAC00" denotes the LO AAC00 parameterization\cite{AAC00}.
GRV98 \cite{GRV98} for unpolarized parton distribution is
used in all the calculations.
}
\label{sigktbag}
\end{figure}

\begin{figure}
\psfig{file=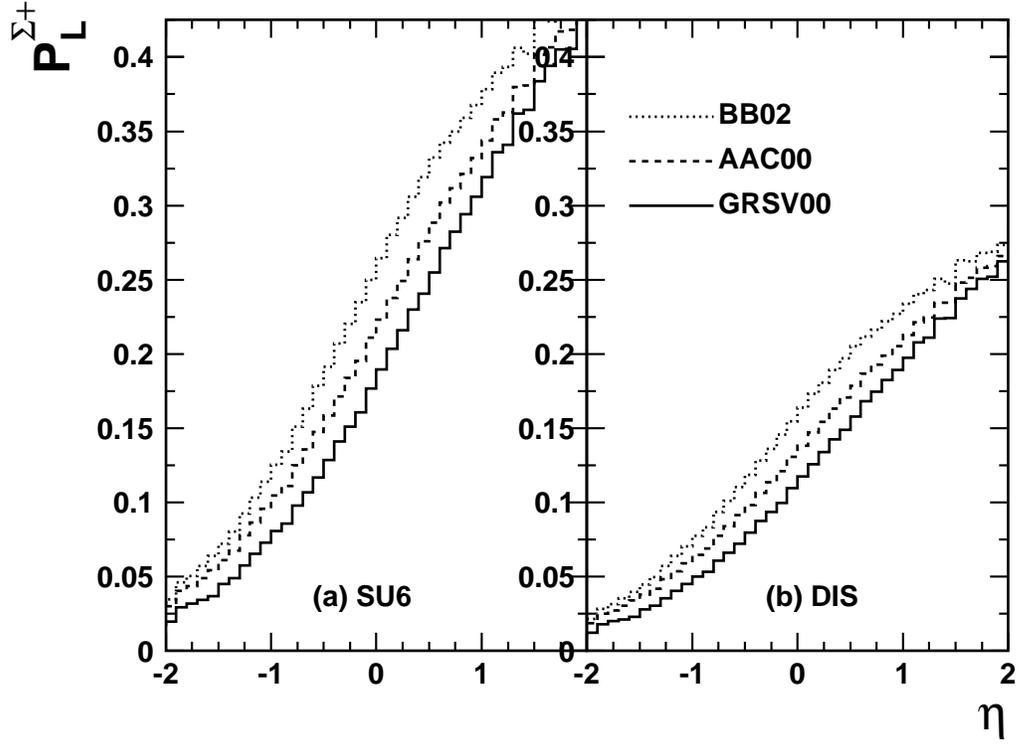,width=16cm}
\caption{
Polarization of $\Sigma^+$
for different sets of $\Delta g(x,\mu^2)$
in $\vec pp \to \gamma \vec{\Sigma}^+ X$ as a function of $\eta$
at $\sqrt s=200$ GeV and $p_T^{min}=8$ GeV
under (a) SU6 picture and (b) DIS picture for the spin transfer factor.
}
\label{sigmutbag}
\end{figure}

Fig. \ref{sigorg} shows the different contributions to $\Sigma^+$
production in $\vec pp\to \gamma \vec \Sigma^+ X$
for $\sqrt s=200$ GeV and $p_T^{min}=4$ GeV.
We see that, the contribution from group (A), i.e.,  
which are directly produced and contain a fragmenting 
$u$ quark is larger than 80\%, even in the small $|\eta|$ region.
This percentage is even
larger than that for $\Sigma^+$ production at high $p_T$ in $pp\to \Sigma^+ X$.
In particular in the large $|\eta|$ region, the $u$ quark fragmentation
plays the dominant role.
Since $u$ quark carries most of the hyperon $\Sigma ^+$'s spin,
the resulting $\Sigma^+$
polarization in $\vec pp \to \gamma {\vec \Sigma} ^+ X$ should
be much larger and more sensitive to $\Delta g(x)$ than that of $\Lambda$.
Furthermore, also due to the dominance of the $u$ quark fragmentation,
the production rate of $\Sigma^+$ should be comparable with
that of $\Lambda$, which implies that the statistics
for studying $\Sigma^+$ should be similar to that of $\Lambda$.
We first calculate $P_L^{\Sigma^+}$ in region of $-1<\eta<0$ as a function of $p_T$ 
at $\sqrt s=200$ GeV under the SU(6) and DIS pictures.
We use three different sets of $\Delta g(x)$ parameterization as inputs
and the results are shown in Fig.\ref{sigktbag}.
The obtained results from different $\Delta g$ parameterizations differ
significantly from each other.
We also show the results of $P_L^{\Sigma^+}$ in Fig.\ref{sigmutbag} for
$p_T^{min}=8$ GeV at $\sqrt s=200$ GeV as a function of $\eta$.
We see that, $P_L^{\Sigma^+}$ increases with increasing $\eta$
to the order of 0.3 at $\eta=1$.
$P_L^{\Sigma^+}$ can select 
among different sets of $\Delta g(x,\mu^2)$ in the region of $\eta<0$.
Such differences could be distinguished at RHIC or other future experiments.
Therefore, by measuring $P_L^{\Sigma^+}$ in high $p_T$ direct
photon production process,
one can get useful information \cite{note1} to distinguish between
different sets of $\Delta g(x,\mu^2)$.
                                                         
As we mentioned in Sec.IIC, the origin of $\Lambda$ inclusively produced
in $pp$ collisions is very complicated and
the decay contribution is very large. 
Also, because of the dominance of $u$ quark fragmentation,
the obtained $\Lambda$
polarization is expected to be very small and contain a large uncertainty.
We do similar calculations and find that the obtained 
$P_L^{\Lambda}$ is smaller than 3\% in all cases and thus it can not be used
to study gluon helicity distribution.
                                                         
\section{Transverse polarization of hyperons with high $p_T$}

When one of the proton beam is transversely polarized, the hyperons
with high $p_T$ can also be transversely polarized.
In this section, we show that
by measuring the polarization of $\Sigma^+$ with high $p_T$ in transversely 
polarized $pp$ collisions, 
one can get useful information on the quark transversity distributions in nucleon.

\subsection{Calculation formulae for hyperon polarization in $\vec p_\perp p
\to \vec H_\perp X$}                                                         

The calculation formulae for hyperon polarization in the transversely
polarized case are completely similar to
those in the longitudinally polarized case.
We summarize them in the following.
Similar to Eq.(\ref{gener1}), the polarization of hyperon
in ${\vec p}_\perp p \to {\vec H}_\perp X$ is given by,
\begin{equation}
P^H_{T}= \frac
{d\sigma^{(p_\uparrow p \to H_\uparrow X)}-d\sigma^{(p_\uparrow p \to
H_\downarrow X)}}
{d\sigma^{(p_\uparrow p \to H_\uparrow X)}+d\sigma^{(p_\uparrow p \to
H_\downarrow X)}}
\equiv \frac
{d\Delta_T \sigma^{(\vec p_\perp p \to \vec H_\perp X)}/d\eta}
{d\sigma^{(pp \to HX)}/d\eta}.
\label{eq1}
\end{equation}
Here, as well as in the following, 
the subscripts ``$\uparrow$" and ``$\downarrow$" denote the 
transverse polarization of the particle.
The polarized cross section has a similar expression as
Eq.(\ref{desig2}),
\begin{equation}
\frac{d\Delta_T \sigma^{(\vec p_\perp p \to \vec H_\perp X)}}{d\eta}
=\int_{p_T^{min}}dp_T
\sum_{abcd}\int dx_a dx_bdz \delta f_a(x_a,\mu^2)f_b(x_b,\mu^2)
 \Delta_T D_c^{H}(z,\mu^2)
 \frac {d\Delta_T \hat {\sigma}^{({\vec a_\perp}b\to {\vec c_\perp} d)}}
{d\eta},
\label{desig}
\end{equation}
where $\Delta_T D_c^H(z,\mu ^2) \equiv
D_{c(\uparrow)}^{H(\uparrow)}(z,\mu
^2)-D_{c(\uparrow)}^{H(\downarrow)}(z,\mu ^2)$
is the transversely polarized fragmentation function; 
$d\Delta_T \hat {\sigma}$ is defined for the hard subprocess 
in case of transverse polarization,
which can also be calculated by pQCD.
Now, similarly to Eq.(\ref{eqaa}), the integrand of Eq.(\ref{desig}) can be
rewritten as,
\begin{equation}
\delta f_a(x_a,\mu^2)f_b(x_b,\mu^2) \Delta_T D_c^H(z,\mu^2)
 {d\Delta_T \hat {\sigma}}
=\sum_{\alpha}P_T^{q_c}T^{F(\alpha)}_{H,c}
f_a(x_a,\mu^2)f_b(x_b,\mu^2)D_c^{H(\alpha)}(z,\mu^2)
{d\hat {\sigma}}.
\label{eq10}
\end{equation}
The only differences are that     
$P_T^{q_c}$ and $T^{F(\alpha)}_{H,c}= {\Delta_T
D_{c}^{H(\alpha)}(z,\mu^2)}/{D_{c}^{H(\alpha)}(z,\mu^2)}$
are the transverse polarization of the scattering quark $c$
and the transverse spin transfer factor,
which take the place of $P_L^{q_c}$ and $t^{F(\alpha)}_{H,c}$
in the longitudinally polarized case.
$P_T^{q_c}$ is related to the transversity distribution by,
\begin{equation}
P_T^{q_c}=D_T
\frac {\delta f_a(x_a,\mu^2)} {f_a(x_a,\mu^2)},
\label{pqT}
\end{equation}
where $D_T\equiv {d\Delta_T \hat {\sigma}}/{d\hat {\sigma}}$ 
is the transverse spin transfer in the hard subprocess 
${\vec a}_\perp b\to  {\vec c}_\perp d $
from the incoming parton $a$ to the outgoing parton $c$. 
To the leading order, it is only a function of $y$ defined in last section,  
and the results for different hard subprocesses can be found in different
publications\cite{SV92,Collins94}.

The product ${f_a(x_a,\mu^2)} f_b(x_b,\mu^2)
{D_c^{H(\alpha)}(z,\mu^2)} {d\hat {\sigma}}$ in the r.h.s. of
Eq.(\ref{eq10}),
is again proportional to the average numbers of
hyperons from different groups (A), (B), (C) and (D) in $pp\to HX$,
which can be calculated by using e.g. {\sc pythia} \cite{PYTHIA}.
The unknown factors left to obtain $P_T^H$
are the spin transfer factor $T^{F(\alpha)}_{H,c}$
and $\delta f_a(x_a,\mu^2)$.
Hence, if we know one of them, one can study the other by measuring $P_T^H$.
 
\subsection{Numerical estimation of transverse polarization of hyperons 
with high $p_T$}

As we emphasized in Sec.II, the relative weights of the different
contributions to hyperons in the final states are determined by the
hadronization mechanism and are independent of the polarization properties.
This means that, if we consider $\vec p_\perp p\to {\vec \Sigma}^+_\perp X$, 
we have the same conclusion that $\Sigma^+$'s of origin (A), i.e.,
those are directly produced and contain a scattered $u$ quark, play the 
dominant role. In this case, for the $\Sigma^+$ polarization, 
the fragmentation effects come in mainly through the transverse spin
transfer factor $T^{F(\alpha)}_{H,c}$, which is a constant.
Similarly, we denote $T^{F(\alpha)}_{H_i,f}$ as the transverse spin
transfer factor for the quark $q_f$ to hyperon $H_i$. 
So, in experiments, if we can determine $T^{F(\alpha)}_{H_i,f}$ using a few
data points, we can extract information on $\delta q(x)$ by measuring
$P_T^{\Sigma^+}$ in pp collisions.

To get a feeling of how large $P_T^{\Sigma^+}$ can be and how strongly it
depends on the different choices of $\delta q(x,Q^2)$,
we make some numerical estimations by using different inputs for
$T^{F(\alpha)}_{H_i,f}$ and $\delta q(x,Q^2)$ in the following. 
In general, the transverse spin transfer factor $T^{F(\alpha)}_{H_i,f}$
can be different from $t^{F(\alpha)}_{H_i,f}$, the spin transfer factor
in the longitudinally polarized case.
This is similar to the difference between the helicity 
and the transversity distributions of the quarks in nucleon \cite{trans02}.
Because of the relativistic effects, the magnitudes and/or shapes of them
are in general different from each other.
On the other hand, it seems that the qualitative features, in particular the
signs of them, are the same, especially in the large momentum fraction
region \cite{trans02}.
Similarly, we may expect that $T^{F(\alpha)}_{H_i,f}$ has the same
qualitative behavior as $t^{F(\alpha)}_{H_i,f}$.
Hence, we use the same results as those for $t^{F(\alpha)}_{H_i,f}$
obtained in the SU(6) and DIS pictures in the following estimations.
For the transversity distribution $\delta q(x,Q^2)$,
we use the simple form obtained
in the light-cone SU(6) quark-spectator model \cite{BQMa,BQMa98},
and for comparison, the upper limit of $\delta q(x,Q^2)$
in Soffer's inequality \cite{Soffer},
\begin{equation}
|\delta q(x,Q^2)| \le \frac{1}{2}[\Delta q(x,Q^2)+q(x,Q^2)].
\end{equation}

\begin{figure}
\psfig{file=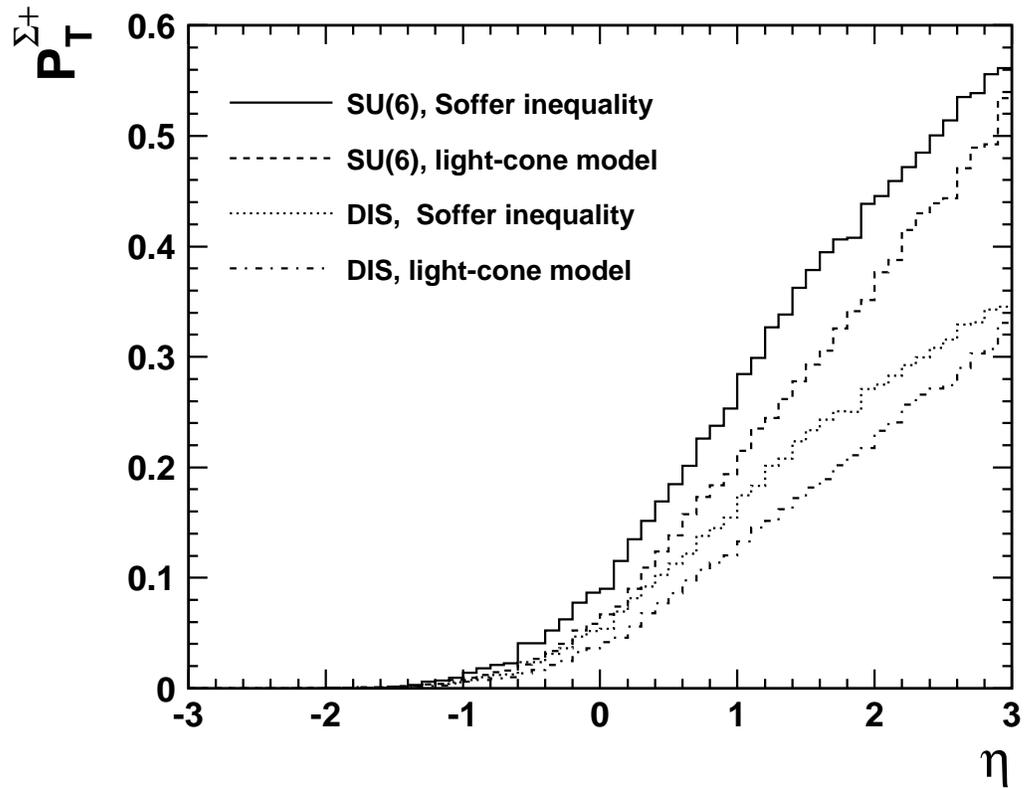,width=16cm}
\caption{$\Sigma^+$ polarization with $p_T^{min}=$13 GeV
in $\vec p_\perp p\to \vec \Sigma^+_\perp X$ as a function of $\eta$
at $\sqrt{s}=$ 500 GeV.
}
\label{sigpol}
\end{figure}

\begin{figure}
\psfig{file=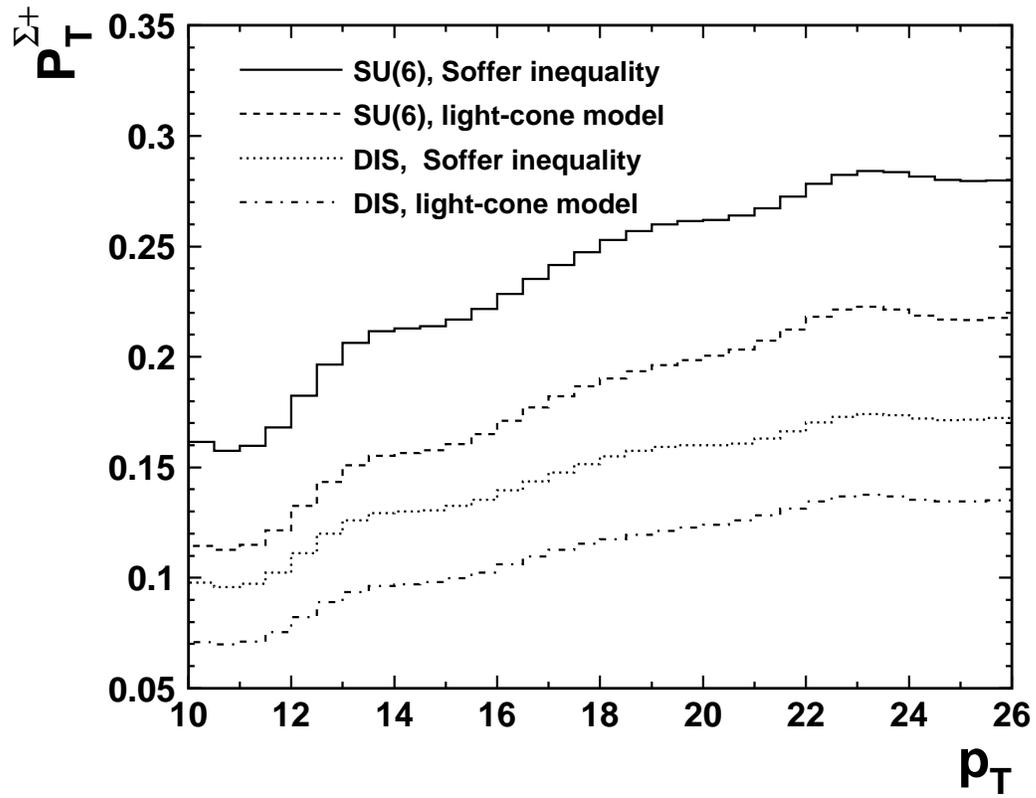,width=16cm}
\caption{$\Sigma^+$ polarization at $0<\eta<1.5$ as
a function of $p_T$
in $\vec p_\perp p\to \vec \Sigma^+_\perp X$ at $\sqrt{s}=$ 500 GeV.
}
\label{sigpolkt}
\end{figure}

We first calculate the transverse polarization of $\Sigma^+$ as a function of
$\eta$ at $\sqrt s=500$ GeV and $p_T^{min}=13$ GeV in $pp$ collisions with one
beam transversely polarized.
Here, the $p_T^{min}$ we have chosen not only guarantees the applicability
of pQCD\cite{Florian98}, but also ensures that the 
quark involved partonic subprocesses dominate  
others\cite{XLL02,Boros00} in $pp\to\Sigma^+ X$.
The results are shown in Fig.\ref{sigpol}.
We see that, the magnitude of $P_T^{\Sigma^+}$ can be quite large.
It increases to about 0.5 with increasing $\eta$.
Though not very large,  
the difference between the Soffer inequality and the light-cone model
is obvious.
In addition, for the convenience of comparison with future experimental data, 
we also calculate $P_{T}^{\Sigma^+}$ for the region of 0$<$$\eta$$<$1.5
as a function of $p_T$ at $\sqrt s=500$ GeV. 
The results are shown in Fig.\ref{sigpolkt}.
We see that, $P_{T}^{\Sigma^+}$'s in different cases increase with $p_T$
and the differences among different models are clear.
$P_{T}^{\Sigma^+}$ could be measured at RHIC or other
future experiments and can be used as a complementary method to obtain $\delta
u(x,Q^2)$.

We should note that, 
the transverse polarization of the produced hyperons discussed above is
defined with respect to the direction of motion of the quark before
fragmentation, or the jet axis in the final state.
It refers to the transverse polarization direction of the quark after
the hard subprocess. This polarization direction is determined by the 
polarization direction of the incoming quark and the scattering process
which is calculable by using pQCD.
The pQCD calculations show that 
the direction of transverse polarization
of the incoming and that of the outgoing quark are related
to each other by a rotation around the normal of the scattering plane,
which changes the moving direction of the quark from
the incoming to the outgoing direction
(C.f. Fig. 2 of Ref.\cite{Collins94}).
In practise, since the hyperons that we consider are mainly the leading
particles in the jet, the jet axis can approximately be replaced by the
direction of motion of the hyperons.

\begin{figure}
\psfig{file=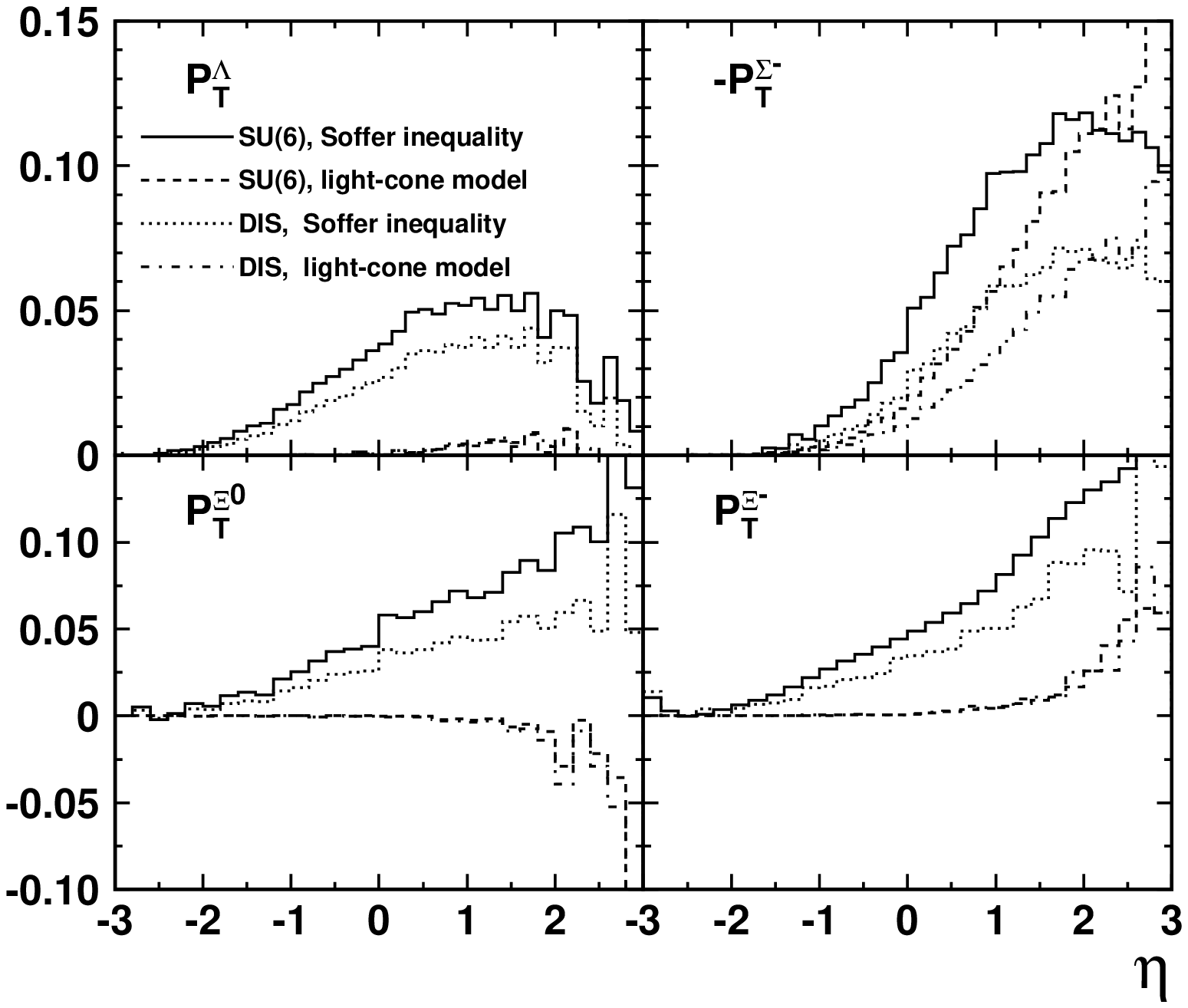,width=16cm}
\caption{Polarizations for $\Lambda$, $\Sigma^-$,
$\Xi^0$ and $\Xi^-$
with $p_T^{min}=$13 GeV
in $\vec p_\perp p\to \vec H_\perp X$ as a function of $\eta$
at $\sqrt{s}=$ 500 GeV.
}
\label{hypol4tt}
\end{figure}

As a comparison, we also calculate the transverse polarizations for 
$\Lambda$, $\Sigma^-$, $\Xi^0$ and $\Xi^-$
with $p_T$$>$13 GeV
in polarized $pp \to H X$ as a function of $\eta$
at $\sqrt{s}=$ 500 GeV.
The results are shown in Fig. \ref{hypol4tt}.
We can see that, they are also transversely polarized with different signs.
$P_{T}^{\Lambda}$ is very small either for
light-cone model or Soffer inequality.
This is because, the spin transfer from the $u$ and $d$ quark to the
produced $\Lambda$ is zero in SU(6) picture and very small in DIS picture
and it is right $u$ and $d$
quark fragmentation that dominate the high $p_T$ hyperon production in $pp$
collisions.

\section{Summary}
In this paper,
we study the polarization of hyperon in different processes in 
polarized $pp$ collisions and its
relation to the polarized parton distributions, in particular
the gluon helicity distribution and the quark transversity distribution.
We show that by
measuring the $\Sigma^+$ polarization in high $p_T$ direct photon
production in singly polarized pp collisions at high energy,
one can extract the gluon helicity distribution;
and present the numerical results for
$\Sigma^+$ polarization in $\vec pp\to \gamma \vec \Sigma^+X$
at RHIC energy obtained using different inputs of $\Delta g(x,Q^2)$.
The results show that
$\Sigma^+$ polarization is in general large and sensitive to 
$\Delta g(x,Q^2)$ in the region of $-2<\eta <0$.
We also find out that the measurement of transverse polarization of $\Sigma^+$ 
with high $p_T$ 
in $\vec p_\perp p\to \vec \Sigma^+_\perp X$ can provide
useful information for the transversity distribution of nucleon,
and present the
numerical results obtained using different inputs for
the transversity distributions and the spin transfer factor
in fragmentation of transversely polarized quark.
Such measurements can be carried out e.g. at RHIC and should provide
useful information on the gluon helicity distribution and quark transversity
distribution in nucleon.

\vspace{0.9cm}
\section*{Acknowledgments}

We thank X.N. Wang for suggesting us to look at the direct photon production
and for stimulating discussions.
This work was supported in part by
the National Science Foundation of China (NSFC) with grant No.10347116 and
No.10175037 and the Education Ministry of China.

\newpage

\begin{thebibliography}{99}
\bibitem{note1}
 We note that the possible NLO corrections
 to the results above are not included here,
 for event generator is used for fragmentation function in our calculations.
 The correction to the spin transfer in the hard
 scattering, the ratio of
 ${d\Delta \hat {\sigma}}/ {d\hat {\sigma}}$, should be
 small \cite{Florian98}.
 So, we expect that such correction to hyperon polarization is small.
 For a precise analysis of the extraction of $\Delta g(x,Q^2)$,
 the NLO corrections should be considered in Eq. (\ref{dirp}).
\end{thebibliography}
\end{document}